\documentclass[letterpaper,12pt]{article}
\usepackage{graphicx}
\usepackage{hhline}
\usepackage{cite}
\usepackage{amsmath,amsfonts}
\newcommand*{\sgn}{\mathop{\rm sgn}\nolimits}
\def\tu{{\bar u}}
\def\tv{{\bar v}}
\def\tx{{\bar x}}
\def\tz{{\bar z}}
\def\tr{{\bar r}}
\def\Tt{{\bar t}}
\def\M{{\cal M}}
\def\O{{\cal O}}
\def\Tr{{\cal T}}

\title{Trap Surface Formation\\ in High-Energy Black Holes Collision}
\author{O.~I.~Vasilenko\thanks{E-mail address:
\texttt{vasilenko@depni.sinp.msu.ru}}
\\[6pt]
\emph{\small Department of Physics, M.V.Lomonosov Moscow State University,}\\
\emph{\small  Moscow 119992, Vorob'evy Gory, Russia}}
\date{}

\begin{document}
\maketitle \abstract{We investigate  classical formation of a trap
surface in  $D$-dimensional Einstein gravity in the process of a
head-on collision of two high-energy particles, which are treated
as Aichelburg-Sexl shock waves. From the condition of the trap
surface volume local maximality we deduce an explicit form of the
inner trap surface. Imposing the continuity condition on the
fronts we obtain a time-dependent solution for the trap surface.
We discuss  trap surface appearance and evolution.

\section{Introduction}\label{introduction}
Now we see a significant interest to  the  processes of  black
hole production in ultra-relativistic particle collisions. A
possibility of such production was shown in a series of papers of
Amati, Ciafaloni and Veneziano
\cite{Amati/Ciafaloni/Veneziano:1987/1989/1989/1990/1992/1993} and
't~Hooft \cite{Hooft:1987/1989/1998}. In
\cite{Aref'eva/Viswanathan/Volovich:9412}, ~Aref'eva, Viswanathan
and Volovich considered a scenario for similar process by using
the Chandrasekhar-Ferrari-Xanthopoulos duality between the Kerr
black hole solution and colliding plane gravitational waves.

A new interest to this problem appeared after the proposition of
Arkani-Hamed, Dimopoulos and Dvali in
\cite{ArkaniHamed/Dimopoulos/Dvali:9803/9807,
Antoniadis/ArkaniHamed/Dimopoulos/Dvali:9804} for solving the
hierarchy problem  based on assumption of existence of additional
spatial dimensions large compared to the weak scale. The
fundamental Planck mass could be of the order of a few TeV which
makes plausible observations of  effects of extra dimensions
existence in future colliders experiments or cosmic rays and
astrophysical measurements (see e.g.
\cite{Mirabelli/Perelstein/Peskin:98118,Banks/Fischler:9906,Aref'eva:9910,
Giddings/Katz:0009,Giddings/Thomas:0106,Giddings:0110/0205/0205/AIP,
Landsberg:0112/0211,Argyres/Dimopoulos/March-Russell:9808,
Dimopoulos/Landsberg:0106,Dimopoulos/Emparan:0108,
Emparan/Horowitz/Myers:0003,Emparan:0302,Ahn/Cavaglia/Olinto:0201,
Ahn/Cavaglia:0205,Hewett/Spiropulu:0205,Mocioiu/Nara/Sarcevic:0301,
Giudice/Rattazzi/Wells:9811/0112, Feng/Shapere:0109,
Anchordoqui/Goldberg:0109/0209,Anchordoqui/Feng/Goldberg/Shapere:0112,
Casadio/Harms:0110,Casadio:0304,Cheung:0110/0205,Ringwald/Tu:0111,Cavaglia:0210,
Voloshin:0107/0111, Hsu:0203, Bilke/Lipartia/Maul:0204,
Chamblin/Nayak:0206,Chamblin/Cooper/Nayak:0301,Kancheli:0208}).

There are some theoretical works
\cite{Cardoso/Lemos:0112/0202/0206/0207,Krasnov:0202,Ida/Oda/Park:0212,
Kanti/March-Russell:0203/0212,Jevicki/Thaler:0203,Solodukhin:0201/0212,
Nakamura/Nakao/Mishima:0302,Eardley/Giddings:0201,Yoshino/Nambu/Tomimatsu:0109,
Yoshino/Nambu:0204,Yoshino/Nambu:0209,Kohlprath/Veneziano:0112/0203}
in which this process was analysed in different aspects.

In \cite{Eardley/Giddings:0201}, Eardley and Giddings developed a
trap surface approach \cite{Penrose:68book,Hawking/Ellis:73book}
in $D$-dimensional Einstein gravity for calculating  the
cross-section of  black hole formation in high-energy collision of
two massless particles, which were treated as Aichelburg-Sexl
shock waves \cite{Aichelburg/Sexl:71,Dray/Hooft:85}. In
\cite{Yoshino/Nambu:0204}, Yoshino and Nambu  investigating the
same task in the case of a head-on collision used the $T=const$
slicing of space-time in the region between the shock-waves to
explore  horizon formation. However, they obtained the surface
that is discontinuous at the shocks and therefore cannot have
trapping properties \cite{Penrose:68book,Hawking/Ellis:73book}.

In this article, we use   a similar to \cite{Yoshino/Nambu:0204}
space-time slicing to consider the problem of trap surface
emergence and evolution. We find a continuous solution for the
surface and investigate it's dynamic.

The paper is organized as follows. We begin in Section~\ref{Shock
waves metric} by discussing singular and nonsingular forms of the
task metric. In Section~\ref{Inner Trap surface}, we formulate an
extremum volume method for trap surface finding and obtain
explicit expressions describing the inner surface in $D$
space-time dimensions. In Section~\ref{Trap surface formation}, we
find  solutions for fronts trap surface and use  continuity
conditions to determine the whole surface. Here, we present
dependencies of surface parameters on time and $D$. Finally, in
Section~\ref{Summaries},  we  summarise and analyse our results.

\section{Shock waves metric}\label{Shock waves metric}

\subparagraph{Coordinate system.} We use a Minkowski coordinate
system $(\Tt,\tz,\tx^i)$ where $\tz$ is the direction of particles
motion and $\tx^i$, $i=1\ldots D-2$ are transverse coordinates.
Let $(\tu,\tv)$ be  light-cone coordinates $(\Tt-\tz,\Tt+\tz)$.
The particles are moving along $\tz$ axis  in the opposite
directions with zero impact parameter ($\tx^i=0$).

\subparagraph{Schwarzschild metric.} In the rest-frame of a sole
isolated particle with mass $M$ metric is described by
D-dimensional Schwarzschild solution
\begin{equation}
ds^2 = -\left(1- \frac{16\pi G_DM}{ (D-2)
\Omega_{D-2}}\frac{1}{R^{D-3}}\right) dt^2 + \left(1- \frac{16\pi
G_DM}{(D-2) \Omega_{D-2}}\frac{1}{R^{D-3}} \right)^{-1} dR^2 + R^2
d\Omega_{D-2}^2\ ,
\end{equation}
where $d\Omega_{D-2}^2$ and $\Omega_{D-2}$ are the line element
and volume of the unit ($D-2$)-sphere
$\left[\Omega_n=2\pi^{(n+1)/2}/\Gamma((n+1)/2)\right]$ and $G_D$
is the $D$-dimensional gravitational constant.

\subparagraph{One shock wave metric.}The gravitational solution
for each of the incoming particles can be found by boosting this
solution, taking the limit of large boost and small mass $M$, with
fixed total energy $\mu$.  The result for a particle moving in the
$+z$ direction is the Aichelburg-Sexl
\cite{Aichelburg/Sexl:71,Dray/Hooft:85} metric
\begin{equation}
\label{Aichelburg-Sexl}
ds^2 = -d\tu d\tv + d\tx^{i2} + \Phi(\tx^i)
\delta(\tu) d\tu^2\ .
\end{equation}
Function $\Phi$ depends only on the transverse radius $\tr=
\sqrt{\tx^i\tx_i}$,  takes the form
\begin{align}
\Phi&= -2a\ln(\tr)\ ,& D=4\ ,\label{Phi4}\\
\Phi&= \frac{2a^{D-3}}{(D-4)\tr^{D-4}}\ ,&D>4\ ,\label{PhiD}
\end{align}
where
\begin{equation}
\label{}a=\left(\frac{8\pi G_D\mu}{\Omega_{D-3}}\right)^{1/(D-3)}
\end{equation}
 and  satisfies the Poisson equation
\begin{equation}
\label{Poisson}
\frac{\partial^2 \Phi}{\partial\tx^i\partial\tx_i} = -16\pi G\mu
\delta^{D-2}(\tx^i)\ .
\end{equation}

\subparagraph{Change of coordinates.} It is possible to remove
singularity in the metric \eqref{Aichelburg-Sexl} by introducing
new coordinates ($u,v,x^i$) defined by
\begin{align}
\tu &= u\ , \notag\\
\tv &= v+\Phi\theta(u) + \frac{u \theta(u) (\nabla\Phi)^2}{4}\ ,\label{Coordinates}\\
\tx^i&= x^i + \frac{u}{2} \nabla_i \Phi(x)\theta(u)\notag
\end{align}
(here $\theta$ is the Heaviside step function). In these
coordinates,  geodesics and their tangents are continuous across
the shock plane at $u=0$.

\subparagraph{Two shock waves metric.} Metric
\eqref{Aichelburg-Sexl} is flat everywhere except  the null plane
$\tu=0$ of the shock wave. So in order to obtain a two shock waves
metric for time $\Tt<0$ preceding the collision  we can combine it
with another similar metric corresponding to the particle moving
along $\tv=0$ in the $-z$ direction by  matching together the
regions of flat space which precede each of two waves.

 In  coordinates \eqref{Coordinates}, the combined metric of the
 shock waves becomes
\begin{gather}
\label{Interval}
 ds^2 = -du\,dv + \left(H^u_{ik}H^u_{jk} + H^v_{ik}H^v_{jk} - \delta_{ij}
\right) dx^i dx^j           \\
\intertext{where}
 H^u_{ij} = \delta_{ij} + \frac{1}{2}\,\frac{\partial^2
\Phi(x)}{\partial x^i\partial x^j}\,u\theta(u)\ ,\quad
H^v_{ij}= \delta_{ij} + \frac{1}{2}\,\frac{\partial^2
\Phi(x)}{\partial x^i\partial x^j} \,v\theta(v)\ . \label{Def-H}\
\end{gather}

 It is convenient to introduce spherical coordinates
$(r,\vec\varphi)\equiv(r,\varphi^k,\ \ k=1,\dots,D-3$) in the
transverse space $(x^i)$.

\section{Inner trap surface}
\label{Inner Trap surface} The event  of  the black hole
production in a two particle collision signifies that during the
collision a space-time region is formed from which  no light ray
can go to infinity. The border of this region is a trap surface.
More precise definition of the trap surface is following
\cite{Penrose:68book}. Trap surface $\Tr$ is a smooth compact
space-like $(D-2)$-surface with the property that light geodesics
which cross it orthogonally converge locally in the future-time
direction. In particular, this signifies that a volume of a
transferred $(D-2)$-surface formed by these geodesics points
placed on a same distance from the trap surface decreases when the
distance increases.

\subparagraph{Slice of space-time.} We shall consider the
following slice of space-time:
\begin{align*}
&{\rm region~I~~~:} \quad t=z,\qquad t\leq T\ ,\\
&{\rm region~II~~:} \quad t=T,\quad T\leq z \leq -T\ ,\\
&{\rm region~III~:} \quad t=-z,\quad t\leq T\ .
\end{align*}
Here $T\leq 0$ and the particles collision takes place at ($T=0,\
z=0$).

\subparagraph{Inner trap surface.} According to this slice,
equations describing a trap surface $\M$\ in the region~II  take
the form
\begin{equation}
\label{InnerTrapEq}
 t=T=const\ , \qquad z=\delta f(r)\ ,\ \delta=\sgn z\ .
\end{equation}
Null normals to this surface are given by
\begin{equation}
\label{InnerTrapNullNormals}
 N(\varepsilon,\delta)=
 \left[N^t,\ N^z,\ N^r,\ \Vec{N^\varphi}\right](\varepsilon,\delta)=
 \left[1,\ -\frac{\varepsilon\delta}{\sqrt{1+f'^2}},\
 \frac{\varepsilon f'}{\sqrt{1+f'^2}},\ \vec 0\right]\ ,
\end{equation}
where $f'=df/dr$ and $\varepsilon=\pm 1$.

A null geodesic which normally crosses the trap surface
\eqref{InnerTrapEq} at ($T,z_0=f(r_0)\equiv f_0,r_0,\vec
\varphi_0$) is a straight line described by equations
\begin{equation}
\label{Geodesic} t=T+\tau,\quad
 z=\delta f_0-\tau\frac{\varepsilon\delta}{\sqrt{1+f'^2_0}},\quad
 r=r_0+\tau\frac{\varepsilon f'_0}{\sqrt{1+f'^2_0}},\quad
\vec \varphi=\vec \varphi_0\ .
\end{equation}

\subparagraph{Transferred surfaces.} Such geodesics transfer the
trap surface $\M$\ on the distance $\tau$ to the surface
$\M(\varepsilon,\tau)$. A volume $S(\varepsilon,\tau)$ of
$\M(\varepsilon,\tau)$ is equal to
\begin{equation}
\label{Volume}
 S(\varepsilon,\tau)=\Omega_{D-3}\int r^{D-3}\,dl\ ,
\end{equation}
where $dl$ is an element of $\M(\varepsilon,\tau)$-surface
generator in ($z,r$)-plane, which according to \eqref{Geodesic}
can be expressed through an initial $dr_0$ as
\begin{equation}
\label{Element}
 dl=\sqrt{dr^2+dz^2}=\sqrt{\left(\frac{dr}{dr_0}\right)^2+
 \left(\frac{dz}{dr_0}\right)^2}\,dr_0=
 \sqrt{1+{f'}_0^2}\left[1+
 \frac{\tau\varepsilon{f'}'_0}{\left(1+{f'}_0^2\right)^{3/2}}\right]\,dr_0\ .
\end{equation}
So \eqref{Volume} can be rewritten as
\begin{equation}
 \label{r_0Integration}
 S(\varepsilon,\tau)=\Omega_{D-3}\int
\left[r_0+\frac{\tau\varepsilon{f'}_0}{\sqrt{1+{f'}_0^2}}\right]^{D-3}\sqrt{1+{f'}_0^2}
 \left[1+\frac{\tau\varepsilon{f'}'_0}{\left(1+{f'}_0^2\right)^{3/2}}\right]
 \,dr_0\ .
\end{equation}
For the small $\tau$ this expression takes the form
\begin{align}
 \label{SmallTauVolume}
 S(\varepsilon,\tau)&=\Omega_{D-3}\int r_0^{D-3}\sqrt{1+{f'}_0^2}
\left\{1+\frac{\tau\varepsilon}{\sqrt{1+{f'}_0^2}}
\left[\frac{(D-3)f'_0}{r_0}+\frac{f''_0}{1+{f'}_0^2}\right]\right.\notag\\
&\left.+
\frac{\tau^2(D-3)}{1+{f'}_0^2}\left[\frac{(D-4)f'^2_0}{2r_0^2}+
\frac{f'_0f''_0}{r_0(1+{f'}_0^2)}\right]+\O(\tau^3)\right\}\,dr_0\
.
\end{align}
\subparagraph{Explicit form of an inner trap surface.} Now we can
derive the equation for function $f(r)$ from the demand that for
the small $\tau$ the volume $S(\varepsilon,\tau)$ decreases when
$\tau$ increases for both $\varepsilon=\pm 1$. The necessary
condition is the equality to zero of the $\tau$-linear term in the
right part of \eqref{SmallTauVolume}
\begin{equation}
 \label{TrapSurfaceEq}
\frac{(D-3)f'_0}{r_0}+\frac{f''_0}{1+{f'}_0^2}=0\ .
\end{equation}
Integration of this equation gives the explicit form of an inner
trap surface
\begin{equation}
 \label{ExplicitTrapSurface}
z=\delta R\int_1^{r/R} \frac{d\rho}{\sqrt{\rho^{2(D-3)}-1}}\ ,
\end{equation}
where $R=R(T)$ is a trap surface radius at $z=0$. For $D=4$ we
have a catenoid
\begin{equation}
 \label{}
r=R\cosh{(z/R)}\ .
\end{equation}
Using \eqref{TrapSurfaceEq}, we can rewrite the
$\tau$-second-order term in \eqref{SmallTauVolume} in the form
\begin{equation}
 \label{Tau^2Term}
\frac{\tau^2(D-3)}{1+{f'}_0^2}\left[\frac{(D-4)f'^2_0}{2r_0^2}+
\frac{f'_0f''_0}{r_0(1+{f'}_0^2)}\right]=
-\,\tau^2\,\frac{(D-2)(D-3)}{2r_0^2}\,\frac{{f'}_0^2}{1+{f'}_0^2}\
,
\end{equation}
which clearly demonstrates that the term is negative and
consequently \eqref{TrapSurfaceEq} is also the sufficient
condition. In fact, inner trap surface calculation is equivalent
to the soap films or minimal surfaces finding task
\cite{Dubrovin/Novikov/Fomenko:1979book}.

\subparagraph{Limitation on $\tau$ range.}Expression
\eqref{r_0Integration} is valid if no self-crossing  of
$\M(\varepsilon,\tau)$-surface takes place. In the case of
$\M(-1,\tau)$, geodesic outgoing from $(z=0,r=R)$-surface must not
reach $(z=0,r=0)$-surface which takes place when $\tau<R$. In the
case of $\M(1,\tau)$, geodesics outgoing from a vicinity of
$(z=0,r=R)$-surface must not cross. Since they cross at $z=0$,
then according to \eqref{Geodesic} $\tau$ cannot exceed
$\tau_{max}$
\begin{equation}
\label{Tau_max} \tau<\tau_{max}=\lim\limits_{r\rightarrow
R}f\sqrt{1+{f'}^2}=\frac{R}{D-3}\ .
\end{equation}
The small $\tau$ limit in \eqref{SmallTauVolume} corresponds to
$\tau\ll R$.

\section{Trap surface formation}
\label{Trap surface formation}

\subparagraph{Front trap surface in the initial coordinates.}From
the point of view of  observers in  outer areas ($|z|>-t$)
collision takes place at
\begin{equation}
\label{OldTrap} \tu=+0\ ,\quad\tv=+0\ .
\end{equation}
Since the metric in these areas is Minkowskian and there is an
axial symmetry, these equations can serve as a definition of  a
trap surface on the shock waves fronts, viz. in the regions I and
II.

\subparagraph{Front trap surface in the new coordinates.} In the
new coordinates ($u,v,r,\varphi^k$), the trap surface in the
region I according to \eqref{Coordinates},\eqref{OldTrap} is given
by
\begin{equation}\label{Trap-u}
u=+0\ ,\quad v+\Phi(r)-\Phi(r_c)=0
\end{equation}
with the null normals $n=\left[n^u,n^v,n^r,\Vec{n^\varphi}\right]$
equal to
\begin{equation}\label{}
n_1=\left[0,~1,~0,~\vec 0\right]\ ,\quad
n_2=\left[\left(\frac{r}{a}\right)^{D-3},~
\left(\frac{a}{r}\right)^{D-3},~1,~\vec 0\right]\ .
\end{equation}
Similarly, the trap surface in the  region II  is described by
\begin{equation}\label{Trap-v}
v=+0\ ,\quad u+\Phi(r)-\Phi(r_c)=0
\end{equation}
and
\begin{equation}\label{}
n_3=\left[1,~0,~0,~\vec 0\right]\ ,\quad
n_4=\left[\left(\frac{a}{r}\right)^{D-3},~
\left(\frac{r}{a}\right)^{D-3},~1,~\vec 0\right]\ .
\end{equation}
Smooth transition from \eqref{Trap-u} to \eqref{Trap-v} demands
$n_2=n_4$ when $u=v=0$. This condition gives the value of the trap
surface radius at the moment of collision $r_c$
\begin{equation}\label{}
r_c\equiv r_b(T=0)=a\ .
\end{equation}

\subparagraph{Trap surface continuity.} Demanding  trap surface
continuity on the border of the regions I and II  and taking into
account \eqref{Trap-u} and \eqref{ExplicitTrapSurface} we obtain
equations for  $R(T)$ and trap surface radius on the border
$r_b(T)$
\begin{equation}\label{Border}
T=\frac{1}{2}\left[\Phi(a)-\Phi(r_b)\right]=-R\int_1^{r_b/R}
\frac{d\rho}{\sqrt{\rho^{2(D-3)}-1}}\ .
\end{equation}
Consideration of the regions II and III border produces the same
equations.

In  \cite{Yoshino/Nambu:0204} where the same problem was treated
one of this condition  was replaced by the demand of null normals
continuity which leads to discontinuity of the surface on the
borders. In our approach, normals $N(1,-1)$ and $n_2$ [$N(1,1)$
and $n_4$] are not equal on the (u=0)[(v=0)-]-border. So in our
case, the surface is continuous but not smooth on the borders.

\subparagraph{Trap surface dynamic.} Dependencies of $R$ on $T$
for $D=4,5,6,7,8,9,10$ are shown in Fig.\ref{fig:fig1}. Trap
surface appears at $t=T_{min}$ when $r_b=r_b(T_{min})$. Further
$r_b$ increases as $T$ increases (Fig.\ref{fig:fig2}) and reaches
its maximum value $a$ at the moment of collision $T=0$.
Dependencies of $T_{min}$, $R(T_{min})$ and $r_b(T_{min})$ on $D$
are given in Table~\ref{Table:MinOnD}.

%
\begin{figure}[h]
\begin{center}
\includegraphics*{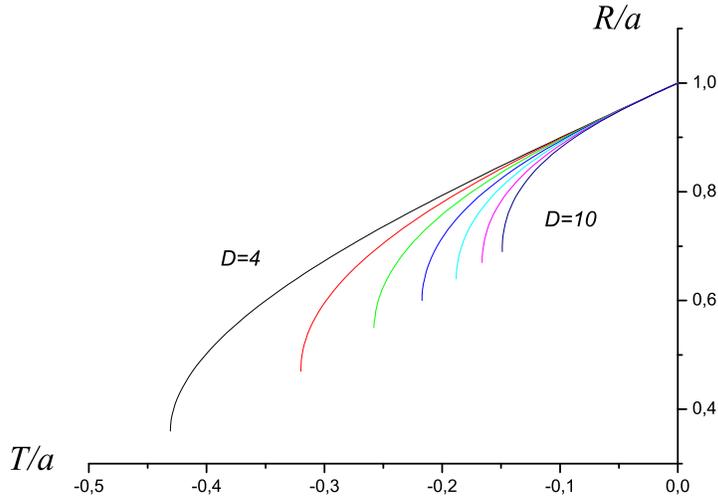}
\caption{\small Dependencies of $R$ on $T$ for $D=4,\dots,10$.}
\label{fig:fig1}
\end{center}
\end{figure}


\begin{table}
\caption{Dependencies of $T_{min}$, $R(T_{min})$, $r_b(T_{min})$
on $D$.}
\label{Table:MinOnD}
\begin{center}
\begin{tabular}{|c|c|c|c|}
\multicolumn{4}{r}{}\\
\hhline{|====|}
 $D$&$T_{min}/a$&$R_{min}/a$&$r_b(T_{min})/a$ \\
\hhline{|=|=|=|=|}
 4&$-0.43$&0.36&0.65\\
\hline 5  & $-0.32$ & 0.47&0.76 \\
\hline 6  & $-0.26$ & 0.54&0.81 \\
\hline 7  & $-0.22$ & 0.60&0.85 \\
\hline 8  & $-0.19$ & 0.64&0.87 \\
\hline 9  & $-0.17$ & 0.67&0.89 \\
\hline 10 & $-0.15$ & 0.69&0.90 \\
\hhline{|=|=|=|=|}
\end{tabular}
\end{center}
\end{table}

%
\begin{figure}[h]
\begin{center}
\includegraphics*{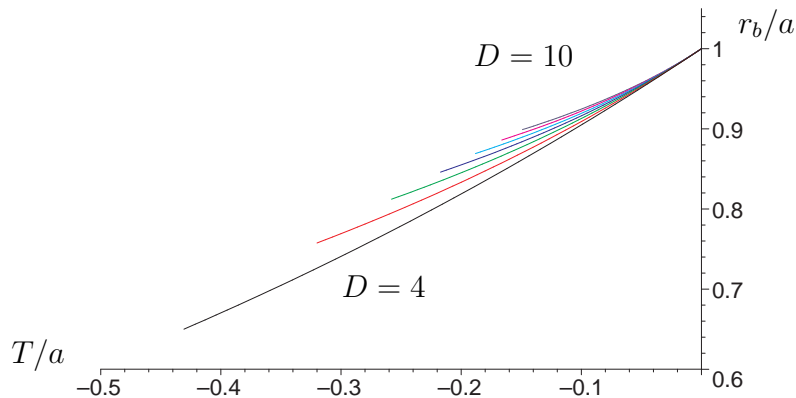}
\caption{\small Dependencies of $r_b$ on $T$ for $D=4,\dots,10$.}
\label{fig:fig2}
\end{center}
\end{figure}

\section{Summaries}
\label{Summaries}
 From the condition of a trap surface volume
local maximality we deduce the explicit form of the inner trap
surface \eqref{ExplicitTrapSurface}. The form of the fronts
surface is given simply by $\tu=+0\ ,\ \tv=+0$ \eqref{OldTrap}.
The condition of  trap surface continuity on the region II borders
\eqref{Border} provides the dependencies on $T$ of $R$ (trap
surface radius at $z=0$) and $r_b$ (trap surface radius at the
shocks) for $D=4,5,6,7,8,9,10$  which are shown in
Fig.\ref{fig:fig1} and Fig.\ref{fig:fig2}.

According to these results, a process of trap surface formation
looks like follows. There is no trap surface at time preceding
$t=T_{min}$. It emerges at $t=T_{min}$ with radius on the shocks
equal to $r_b(T_{min})$. Values of $T_{min}$, $R(T_{min})$ and
$r_b(T_{min})$ grow with a raise of $D$ (see
Table~\ref{Table:MinOnD}). Later $r_b$ and $R$ increase with time
and reach their maximum value $a$ at the moment of collision.

There are many speculations predicting the end of short distance
physics.  Calculation of the size of horizon formation in black
hole collisions may help us to evaluate a distance at which this
end may take place.

\section*{Acknowledgments}
Conversations with I.~Ya.~Aref'eva are gratefully acknowledged.

\end{document}